\documentclass[manuscript,screen]{acmart}

\AtBeginDocument{%
  \providecommand\BibTeX{{%
    \normalfont B\kern-0.5em{\scshape i\kern-0.25em b}\kern-0.8em\TeX}}}

\begin{document}

\title[The Personalization Paradox]{The Personalization Paradox: the Conflict between Accurate User Models and Personalized Adaptive Systems}


\author{Santiago Onta\~{n}\'{o}n}\authornote{Currently at Google.}
\affiliation{%
  \institution{Drexel University}
  \city{Philadelphia, PA}
  \country{USA}}
\email{santi.ontanon@gmail.com}

\author{Jichen Zhu}\authornote{Currently at the IT University of Copenhagen, Denmark.}
\affiliation{%
  \institution{Drexel University}
  \city{Philadelphia, PA}
  \country{USA}}
\email{jichen.zhu@gmail.com}

\renewcommand{\shortauthors}{Zhu and Onta\~{n}\'{o}n}

\begin{abstract}
Personalized adaptation technology has been adopted in a wide range of digital applications such as health, training and education, e-commerce and entertainment. Personalization systems typically build a {\em user model}, aiming to characterize the user at hand, and then use this model to {\em personalize} the interaction. Personalization and user modeling, however, are often intrinsically at odds with each other (a fact some times referred to as the {\em personalization paradox}). In this paper, we take a closer look at this personalization paradox, and identify two ways in which it might manifest: {\em feedback loops} and {\em moving targets}. To illustrate these issues, we report results in the domain of personalized exergames (videogames for physical exercise), and describe our early steps to address some of the issues arisen by the personalization paradox.
\end{abstract}



\maketitle

\section{Introduction}

{\em Personalized adaptation} is the idea of designing systems that automatically adapt the user experience to the current user. 
This idea has been adopted in a wide range of digital applications such as health, training and education, e-commerce, and entertainment~\cite{zhu2020Personalization}. By using artificial intelligence (AI) to tailor themselves to individual users' needs and preferences, adaptive digital applications have shown to improve learnability~\cite{furqan2017learnability}, usability~\cite{hook1998evaluating} and user enjoyment~\cite{shaker2010towards}. Moreover, personalization technology has also risen a number of open challenges (such as interpretability or controllability~\cite{jameson2007adaptive}). 
This paper focuses on an inherent problem of personalized systems which we refer to as the {\em personalization paradox}~\footnote{The term ``personalization paradox'' is some times also used to refer to the ``privacy-personalization paradox''~\cite{aguirre2015unraveling} (personalization creates users' sense of vulnerability and lower adoption rates), which is different from the use in this paper.}. The term was first used by Jarno Koponen, who, in a 2015 {\em TechCrunch} article\footnote{Retrieved from \url{https://techcrunch.com/2015/06/25/the-future-of-algorithmic-personalization/}, Oct 1, 2020.}, described it as: 
\begin{quote}
``{\em there lies a more general paradox at the very heart of personalization. Personalization promises to modify your digital experience based on your personal interests and preferences. Simultaneously, personalization is used to shape you, to influence you and guide your everyday choices and actions. Inaccessible and incomprehensible algorithms make autonomous decisions on your behalf. They reduce the amount of visible choices, thus restricting your personal agency.}''
\end{quote}
While this is an insightful observation, the discussion around the personalization paradox has so far remained abstract. Reflecting on existing work on personalization, we argue that the personalization paradox is a result of the fundamental {\bf conflict between user modeling and personalized adaptation}. We use this angle to analyze the paradox deeper and identify two main ways in which the paradox can manifest when integrating user modeling and personalization, namely: {\em feedback loops} and {\em moving targets}.

In the remainder of this paper, we first briefly introduce some background on user modeling and personalization, then present our analysis of the personalization paradox, and conclude by illustrating the issues we identified within the context of a personalized system in the domain of exergames~\cite{caro2018understanding}, aiming at behavioral change to promote physical activity. For more details on our work on personalized exergames, and how the personalization challenge arose in our experiments, the reader is referred to~\cite{zhu2021cscw}.

\section{Background}

A significant amount of research exists in personalized adaptation, especially in Internet applications~\cite{Churchill2013}. This ranges from just offering customization options that users themselves configure to complex automatic customization using machine-learned models of user preferences, needs and behavior~\cite{Kramer2000}. These systems are typically composed of two subsystems~\cite{jameson2007adaptive,zhu2020Personalization}: {\em user modeling} and {\em adaptation} (also referred to as {\em user model acquisition} and {\em user model application}).

User modeling approaches range from the collection of user profiles~\cite{gauch2007user} to the use of machine learning algorithms~\cite{li2020survey}, and has been applied to domains such as e-commerce~\cite{goy2007personalization}, training/education~\cite{kass1989student}, social media (e.g. Facebook, Twitter), health~\cite{chawla2013bringing}, and video games~\cite{yannakakis2013player}. Depending on the application domain, user models has been used to capture a range of user characteristics. For example, in the domain of video games, work exists on capturing player behavior~\cite{Drachen2009,valls2015exploring,holmgaard2014evolving,harrison2011using}, player types~\cite{bartle1996hearts,heeter2011beyond}, user preferences~\cite{thue2008passage,sharma2010drama}, skill level~\cite{missura2009player,jennings2010polymorph,zook2012temporal}, knowledge~\cite{kantharaju2019tracing} and goals~\cite{ha2011goal}. 

User models are then used to adapt the user experience given a target {\em personalization goal}. This personalization goal depends on the target application. For example, in a personalized exergame, the goal is to personalize the experience to maximally motivate the user to perform physical activity~\cite{zhu2021cscw}. 
In training/education, the goal is to maximize learning outcomes~\cite{kass1989student, kantharaju2019tracing}. In summary, a standard personalization system works by building a user model, and then adapting the experience based on that model, attempting to affect user behavior in order to achieve the personalization goals.

\section{The Personalization Paradox}

We argue that the personalization paradox is a result of the fundamental conflict between user modeling and personalized adaptation. The key underlying problem is that while user modeling tries to acquire a model of some aspects of interest of the user (such as their preferences), personalized adaptation changes the context the user interacts with. There is a large body of literature in psychology that consider human behavior and preferences to be contextual, hence pointing out that that personalization influences the user herself. This ``influencing the subject that we are trying to model'' issue can manifest in several ways. Through an analysis of existing work, we identified two main forms this paradox can take, listed below and illustrated in Figure \ref{fig:paradox} (notice these might not be mutually exclusive):

\begin{figure}[t!]
\includegraphics[width=0.8\columnwidth]{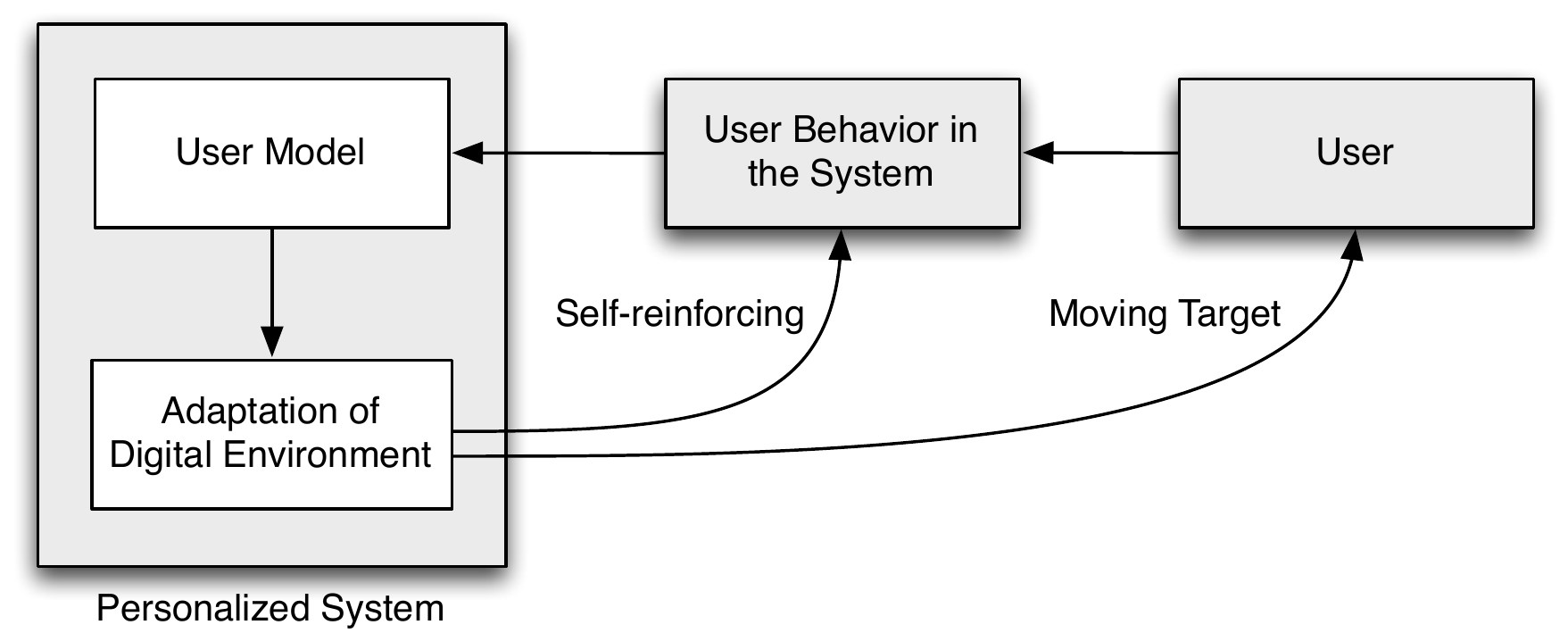}
\centering
\caption{Personalization might create (1) Self-reinforcing loops that push user behavior to become what the system predicts, and (2) Make the user a moving target for user modeling by changing the digital environment context.}
\label{fig:paradox}
\end{figure}

{\bf Self-Reinforcing Loops}: This problem happens when a personalized system ``forces'' a user into what its user model categorizes, regardless whether the model is accurate. This self-reinforcing nature of personalized technology has been documented by other researchers~\cite{o2016weapons,noble2018algorithms,pariser2011filter}. For instance, imagine the {\em Netflix}'s user model inaccurately predicts a specific user's preference to be only Sci-fi based on her viewing history and thus only recommends Sci-fi content. Since the user can only express her preference through the digital environment controlled by the personalization algorithm, she is more likely to further display Sci-fi preference due to the lack of other choices. In this way, the adaptation reinforces its user model without a chance to adjust the latter. 
Another recent example was reported in a recent study by Arnold et al.~\cite{arnold2020predictive}, who showed that adaptive predictive text interfaces affect the text being written by users (in particular, it encourages predictable and shorter writing). As the authors put it ``predictive text encourages predictable writing''.

{\bf Moving Target}: Even when no self-reinforcing loops occur, the fact that personalization affects the user's behavior is still problematic. 
Continuing the above example, let us assume that the user model correctly categorized a user's preference to Sci-fi when she is in the ``neutral'' viewing environment with a wide range of different genres. However, when adaptation personalizes this context into a Sci-fi-heavy one, the user's preference may change to, say, documentary. This case is the closest to Kopenen's description above --- by changing the users' digital environment through adaptation, users' preferences and behavior become a moving target for modeling. A special case of this problem occurs when the goal of the personalization system is to induce behavior change. In this situation, the system's explicit goal is to push the user's preferences or behavior in a particular direction. As a result, user modeling might reflect the user at the start, rather than what she has become.

\subsection{The Personalization Paradox in Adaptive Exergames}

In our recent work on personalization in the context of physical activity (PA)~\cite{gray2020player}, we designed an adaptive system based on {\em social comparison}~\cite{arigo2020social} to motivate users to perform PA. Specifically, we designed a web-based platform in which users can compare themselves, including their daily steps with other users' PA-related profiles. The user's steps are captured by {\em Fitbit} and synced automatically with our platform. We use the AI technique of multi-armed bandits to model individual users' social comparison preferences and adapt the comparison targets shown to them~\cite{gray2020player,gray2020cog}. For example, if the user model predicts that a particular user tends to prefer {\em upward comparisons}, the system will show more profiles with a larger number of daily steps.

Concerning feedback loops, in our work, we hypothesize that when the users become more physically active, they will perform less downward comparison (when they compare themselves with other users that performed less physical activity). What is captured in the user model is often connected to the desired actions by design. In our case, our underlying assumption is that whose activity level a user compare herself to (upwards or downwards) will impact the user' PA and motivation. Thus, if were to add other mechanics to our system that further foster these changes (e.g., competition), this will likely influence players towards upward comparison behaviors, creating a feedback loop, and thus affecting user modeling. In order to avoid this self-reinforcing loop, we avoided the inclusion of such mechanics, and also provided different profiles the users could choose to compare themselves to. This was designed this way to allow them to express a comparison preference different from what the user model predicted at the time. 

Our attempt to minimize the {\em moving target problem}, however, was met with mixed success. In our domain, the personalization component affects the type of other user profiles the user can choose for comparison. Similarly as the NetFlix example above, as the personalization component changes the distribution of profiles, it is unclear if the user behavior will remain constant. 
For instance, when a user increases her PA, will she compare upwards more often? Currently, psychology literature does not have a definitive answer. If so, the moving target problem of the changing user in our case may be exacerbated by the targeted behavior change, in addition to the adapted digital environment.

To mitigate this Moving Target problem, we used two strategies. First, our user modeling does not try to model user behavior based on their comparison choice, but models the expected amount of physical activity a user will be expected to do in each of the contexts our personalization system can choose from. Second, we designed the app to be as neutral as possible --- we intentionally left out the design elements (e.g., competition) which are known to increase engagement and PA but risk pushing users towards upward comparison, penalizing downward comparison. However, this may have significantly reduced the app's ability to motivate changes in PA.  Conversely, if we had decided to incorporate features known to motivate PA effectively, we risk skewing users' preferences and jeopardizing the accuracy of the user model.


\section{Discussion}

Our project shows that the tension created by the personalization paradox is especially prominent when combining personalization with behavioral change. To accurately capture and model users' innate social comparison preferences and reactions, we designed the app to be as neutral as possible. By removing design elements such as competition and goals, we aimed to avoid incentivizing upward comparison and penalize downward comparison, and thus to allow users to express their ``innate'' preference for social comparison. However, this may have significantly reduced the app's ability to motivate changes in PA.  Conversely, if we had decided to incorporate features known to motivate PA effectively, we risk skewing users' preferences and thus jeopardizing the foundation of personalization. For more details on these experiments, the reader is referred to~\cite{zhu2021cscw}.

To mitigate the personalization paradox, we believe that the first step is for the AI engineers/data scientists and designers/behavioral scientists to have an open conversation about the priorities of the project. In our exergame use-case, for example, given the lack of work in modeling social comparison in exergames, prioritizing player modeling accuracy was appropriate. 
Another approach that is worth studying is to further separate the user modeling and adaptation stages of personalization. For example, our app could first, during a period of time, stay neutral as it is in our study to collect accurate user data on their social comparison tendencies. Once it has a robust model, then the app could unlock other design features to explicitly motivate PA. However, more research is needed to see if users' social comparison tendencies modeled in a neutral environment can be transferred into a different context with new features that may reward upward comparison. As this research matures, we may want to strike a different balance to have stronger behavioral outcomes. As personalization becomes increasingly used in areas associated with behavior change further research is needed to balance personalization model accuracy and behavioral change effectiveness.

\begin{acks}
This work is partially supported by the National Science Foundation (NSF) under Grant Number IIS-1816470. The authors would like to thank all past and current members of the project.
\end{acks}

\bibliographystyle{ACM-Reference-Format}


\end{document}